# Cooperative Guidance Strategy for Active Defense Spacecraft with Imperfect Information via Deep Reinforcement Learning

Zhi Li, Haizhao Liang, Jinze Wu, and Jianying Wang.
*Sun Yat-sen University, Guangzhou, 510275, China*

Yu Zheng
*Science and Technology on Space Physics Laboratory, Beijing, 100076, China*

**In this paper, an adaptive cooperative guidance strategy for the active protection of a target spacecraft trying to evade an interceptor was developed. The target spacecraft performs evasive maneuvers, launching an active defense vehicle to divert the interceptor. Instead of classical strategies, which are based on optimal control or differential game theory, the problem was solved by using deep reinforcement learning method and imperfect information was assumed for the interceptor maneuverability. To address the sparse reward problem, a universal reward design method and an increasingly difficult training approach were presented utilizing the shaping technique. Guidance law, reward function, and training approach were demonstrated through the learning process and Monte-Carlo simulations. The application of the non-sparse reward function and increasingly difficult training approach accelerated the model convergence, alleviating the overfitting problem. Considering a standard optimal guidance law as a benchmark, the effectiveness and advantages, that guaranteeing the target spacecraft's escape and win rates in multi-agent game, of the proposed guidance strategy were validated by the simulation results. The trained agent adaptiveness to the interceptor maneuverability was superior to the optimal guidance law. Moreover, compared to the standard optimal guidance law, the proposed guidance strategy performed better with less prior knowledge.**

## Nomenclature

| | | |
|---|---|---|
| $a$ | = | acceleration, m/s² |
| $ac$ | = | action of the agent |
| $\mathbf{A}, \mathbf{B}$ | = | state-space model of the linearized equations of motion |
| $H$ | = | Hamiltonian |
| $\mathbf{I}$ | = | identity matrix |
| $J(\cdot)$ | = | cost function |
| $\mathbf{L}$ | = | constant vector |
| $L^{-1}$ | = | inverse Laplace transform |
| LOS | = | light-of-sight |
| $Q(\cdot)$ | = | Q-value function |
| $r$ | = | reward signal |
| $s$ | = | state defined in Markov decision process |
| $t, t_{go}, t_f$ | = | time, time to go, and final time, respectively, s |
| $u$ | = | guidance command, m/s² |
| $v$ | = | velocity, m/s |
| $X-O-Y$ | = | Cartesian reference frame |
| $\mathbf{x}$ | = | state vector of the linearized equations of motion |
| $y$ | = | lateral distance, m |
| $Z$ | = | zero-effort-miss, m |
| $\alpha, \beta, \sigma$ | = | design parameters of the reward function |
| $\phi$ | = | flight path angle, rad |
| $\mathbf{\Phi}$ | = | transition matrix |
| $\gamma$ | = | discount factor |
| $\eta$ | = | killing radius, m |
| $\lambda$ | = | the angle between the corresponding light-of-sight and X-axis, rad |
| $\lambda(\cdot)$ | = | Lagrange multiplier vector |

| $\mu(\cdot)$ | = | policy function |
| $\rho$ | = | relative distance between the adversaries, m |
| $\tau$ | = | time constant |
| $\omega$, $\xi$ | = | design parameters of the optimal guidance law (OGL) |

*Subscripts*

| I, T, D | = | interceptor, target, and defender, respectively |

*Superscripts*

| max | = | maximum |
| * | = | optimal solution |

## I. Introduction

During the decades, thousands of launching tasks have make space crowded. Spacecraft such as satellites, space stations, and space shuttles, playing an important role in both civil and military activity, are also under directly hit-to-kill threat from similar vehicles or other kind of interceptors. In an effort to avoid the interception, the spacecraft can launch an active-defense vehicle (ADV) as a defender to intercept the incoming threat, which has been widely researched. This countermeasure was proven to be an effective approach to compensate for the inferior target maneuverability [1-3].

A number of guidance laws such as optimal guidance laws [1, 4, 5], differential game guidance laws [2, 3, 6, 7] and nonlinear guidance laws [8], were developed for this active-defense scenario participants. Optimal control-based guidance laws assume that the adversary's maneuvering is known in advance. In [1], optimal cooperative pursuit and evasion strategies were proposed by using Ponrtyagin's minimum principle. A similar scenario was investigated in [2] for both the continuous and discrete domain by using the linear-quadratic differential game method. It is worth noting that the differential game guidance strategies proposed in [2] solve the fuel cost and saturation problem. However, differential game guidance laws suffer from the distinct disadvantages because of the complex calculations and parameter tuning. A switching surface [3], designed using zero-effort miss distance, was introduced to split the multi-agent engagement into two one-on-one differential games, making a trade-off between the performance and efficiency. Even so, it currently does not appear to be implementable in real time.

In recent years, several works have demonstrated improved performance with real-time guidance requirement and complex dynamics by incorporating the traditional guidance method and machine learning [9-11]. In [9], the authors proposed a novel deep neural network (DNN) based online predictive guidance algorithm by combining the existing predictive guidance and the neural network predictor. In [10], an intelligent predictor-corrector entry guidance approach for the high lifting vehicles was proposed by leveraging the DNN and constraint management techniques. In general, the DNN-based guidance strategies generally include two steps. The first step includes generating optimal trajectory data set by using analytical strategies. In the second step, the pre-generated trajectories are then used to train the DNN, which acts as the optimal real-time command generator. However, the aforementioned two-step strategy is only suitable for the deterministic scenario, where it is easy to generate the optimal trajectory. Unfortunately, it is difficult to obtain abundant training data needed for the active-defense scenario by using existing guidance strategies, mostly due to challenges arising from the parameter choice. Moreover, the optimal solution is unavailable if the information is noise-corrupted and real-time performance is required.

A different guidance approach is represented by deep reinforcement learning (DRL), which was successfully applied in a wide range of decision-making problems; examples include robotic control, MOBA game, autonomous driving, and navigation [12-16]. In [17], RL was used to learn an adaptive homing-phase guidance law considering both the sensor and actuator noises and delays. Thus, RL has the advantage over optimal control theory since it can learn using high-fidelity spacecraft and environment models. In [18], an adaptive guidance system designed to solve the landing problem was proposed. The reinforcement meta-learning was used in this approach, and the agent training under a specific environment adapted to a new environment with limited training steps. This has shown that the ability to utilize parameter uncertainty leads to robust policies. A system that can map imperfect information directly to guidance commands has great potential advantages. However, unlike the single-agent cases, where the cost function (rewards in the reinforcement learning framework) is intuitive, the reward design of multi-agent games may encounter several problems. First, the reward signal is extremely sparse. the reward signal is extremely sparsethe players generally have contradictory tasks, making the process more difficult. Thus, it is necessary to design an adaptive reward function to promote cooperation among the agents.

Motivated by the previously presented overview, the authors studied the multi-agent pursuit-evasion scenario with DRL. The solution was meant to have real-time performance, greater adaptiveness, and better performance in the absence of perfect information. Firstly, the kinematic and dynamic models of the engagement scenario are derived.

The agent-environment interface was constructed using a RL algorithm and engagement dynamics. Additionally, OGLs were given as a benchmark to enable the performance comparison, as well as comparing the application requirements. Furthermore, a specific reward function and increasingly difficult learning schemes created by employing shaping techniques were provided. Finally, the learning process and Monte-Carlo simulations were presented aiming to validate the effectiveness of the proposed approach.

The remainder of this paper is organized as follows: in Section II, the authors formulated the multi-agent pursuit-evasion game. In Section III, the analytical analysis of the proposed guidance strategy was presented, while the simulation analysis was given in Section IV. In the same section, the proposed method was compared to its analytical counterpart.

## II. Problem Formulation

A multi-agent game is considered, which has a spacecraft as the main target (T), an active protection vehicle as a defender (D), and a high agility small spacecraft as an interceptor (I). In this engagement, the interceptor is chasing the target, which launches the defender to protect itself by destroying the interceptor. During the endgame, all the players are considered to be constant-speed mass points trajectories of which can be linearized around the initial line-of-sight. As a trajectory linearization consequence, the engagement, a three-dimensional process, can be simplified and will be analyzed in a plane. However, it should be taken into account that in most cases those hypotheses will not harm the result generality [19].

A schematic view of the engagement is shown in Fig. 1, where $X-O-Y$ is a Cartesian inertial reference frame, $\rho$ represents the relative distance, $y$ represents the altitude, $V$ represents the velocity, $a$ represents the acceleration and $\phi$ represents the flight path angle, respectively. Variables are associated with the target, defender, or interceptor, as indicated by subscripts T, D, and I, respectively. The lines-of-sight between the adversaries are denoted by $LOS_{IT}$ and $LOS_{ID}$, while the angles between the line-of-sight and X-axis are defined by $\lambda_{IT}$ and $\lambda_{ID}$. Finally, the relative distance components perpendicular to X-axis are referred to as $y_{IT}$ and $y_{ID}$.

Bearing in mind the mission of all the players, the target is required to evade the interceptor with the help of the defender. At the same time, the interceptor is required to evade the defender and pursuit the target. Thus, the target pair guidance law will attempt to converge $y_{ID}$ to zero and to maximize $y_{IT}$. For the interceptor, the guidance law has the opposite role (its geometric meaning is shown in Fig. 1). The scenario can thus be divided into two collision triangles; the first is between the interceptor and the target, while the second is between the interceptor and the defender.

## A. Equations of Motion

Consider the I-T collision triangle and the I-D collision triangle in a multi-agent pursuit-evasion engagement. The kinematics is expressed using the polar coordinate system attached in the target and defender as follows:

$$\begin{aligned}
\dot{\rho}_{IT} &= -V_I \cos(\phi_I + \lambda_{IT}) - V_T \cos(\phi_T - \lambda_{IT}) \\
\dot{y}_{IT} &= V_I \sin\phi_I - V_T \sin\phi_T \\
\dot{\lambda}_{IT} &= \frac{V_I \sin(\phi_I + \lambda_{IT}) - V_T \sin(\phi_T - \lambda_{IT})}{\rho_{IT}}
\end{aligned} \quad (1)$$

$$\begin{aligned}
\dot{\rho}_{ID} &= -V_I \cos(\phi_I + \lambda_{ID}) - V_D \cos(\phi_D - \lambda_{ID}) \\
\dot{y}_{ID} &= V_I \sin\phi_I - V_D \sin\phi_D \\
\dot{\lambda}_{ID} &= \frac{V_I \sin(\phi_I + \lambda_{ID}) - V_D \sin(\phi_D - \lambda_{ID})}{\rho_{ID}}
\end{aligned} \quad (2)$$

Furthermore, the flight path angles associated with dynamics can be defined for each of the players:

$$\dot{\phi}_i = \frac{a_i}{V_i}, \quad i = \{\text{I, T, D}\} \quad (3)$$

## B. Linearized Equations of Motion

By applying the initial trajectory linearization assumption, the equations of motion can be linearized around the initial line-of-sight:

$$\begin{aligned}
\dot{\rho}_{IT} &= -(V_I + V_T) \\
\dot{y}_{IT} &= a_I - a_T \\
\dot{\lambda}_{IT} &= 0 \\
\dot{V}_I &= 0 \\
\dot{V}_T &= 0
\end{aligned} \quad (4)$$

$$\begin{aligned}
\dot{\rho}_{ID} &= -(V_I + V_D) \\
\dot{y}_{ID} &= a_I - a_D \\
\dot{\lambda}_{ID} &= 0 \\
\dot{V}_I &= 0 \\
\dot{V}_D &= 0
\end{aligned} \quad (5)$$

The dynamics for each of the players is assumed to be a first-order process:

$$\dot{a}_i = -\frac{a_i - u_i}{\tau_i}, \quad i = \{\text{I, T, D}\} \tag{6}$$

where $\tau_i$ is the time constant of the players' dynamics and $u_i$ is the guidance command bounded by $u_i^{\max}$.

Furthermore, the variable vector can be defined as follows:

$$\boldsymbol{x} = \begin{bmatrix} y_{\text{IT}} & \dot{y}_{\text{IT}} & y_{\text{ID}} & \dot{y}_{\text{ID}} & a_{\text{I}} & a_{\text{T}} & a_{\text{D}} \end{bmatrix} \tag{7}$$

while the linearized equations of motion in the state space form can be written as:

$$\dot{\boldsymbol{x}} = \boldsymbol{A}\boldsymbol{x} + \boldsymbol{B}\begin{bmatrix} u_{\text{I}} & u_{\text{T}} & u_{\text{D}} \end{bmatrix} \tag{8}$$

where:

$$\boldsymbol{A} = \begin{bmatrix} 0 & 1 & 0 & 0 & 0 & 0 & 0 \\ 0 & 0 & 0 & 0 & 1 & -1 & 0 \\ 0 & 0 & 0 & 1 & 0 & 0 & 0 \\ 0 & 0 & 0 & 0 & 1 & 0 & -1 \\ 0 & 0 & 0 & 0 & -1/\tau_{\text{I}} & 0 & 0 \\ 0 & 0 & 0 & 0 & 0 & -1/\tau_{\text{T}} & 0 \\ 0 & 0 & 0 & 0 & 0 & 0 & -1/\tau_{\text{D}} \end{bmatrix} \tag{9a}$$

$$\boldsymbol{B} = \begin{bmatrix} 0_{5\times 3} \\ \boldsymbol{B}_1 \end{bmatrix}, \quad \boldsymbol{B}_1 = \begin{bmatrix} 1/\tau_{\text{I}} & 0 & 0 \\ 0 & 1/\tau_{\text{T}} & 0 \\ 0 & 0 & 1/\tau_{\text{D}} \end{bmatrix} \tag{9b}$$

Since the velocity of each player is assumed to be constant, the engagement can be formulated as a fixed-time process. Thus, the interception time can be calculated using:

$$\begin{aligned} t_{f\text{IT}} &= -\rho_{\text{IT}}^0/\dot{\rho}_{\text{IT}} = \rho_{\text{IT}}^0/(V_{\text{I}} + V_{\text{T}}) \\ t_{f\text{ID}} &= -\rho_{\text{ID}}^0/\dot{\rho}_{\text{ID}} = \rho_{\text{ID}}^0/(V_{\text{I}} + V_{\text{D}}) \end{aligned} \tag{10}$$

where $\rho_{\text{IT}}^0$ represents the initial relative distance between the interceptor and the target, while $\rho_{\text{ID}}^0$ is the distance between the interceptor and the defender, allowing to define the time to go by

$$\begin{aligned} t_{\text{goIT}} &= t_{f\text{IT}} - t \\ t_{\text{goID}} &= t_{f\text{ID}} - t \end{aligned} \tag{11}$$

**C. Zero-Effort Miss**

A well-known zero-effort miss (ZEM) was introduced in the guidance law design and reward function design. It is obtained from the homogeneous solutions of equations of motion and is only affected by the current state and interception time. It can be calculated as follows:

$$Z_{IT}(t) = L_1 \Phi(t, t_{fIT}) x(t)$$
$$Z_{ID}(t) = L_2 \Phi(t, t_{fID}) x(t) \tag{12}$$

where

$$L_1 = [1 \ 0 \ 0 \ 0 \ 0 \ 0 \ 0]$$
$$L_2 = [0 \ 0 \ 1 \ 0 \ 0 \ 0 \ 0] \tag{13}$$

and $\Phi$ is the dynamic transition matrix. Thus, the ZEM and its derivative with respect to time are given as:

$$Z_{IT}(t) = x_1 + t_{goIT} x_2 + a_I \tau_I^2 \varphi(t_{goIT}/\tau_I) x_5 - a_T \tau_T^2 \varphi(t_{goIT}/\tau_T) x_6$$
$$Z_{ID}(t) = x_3 + t_{goID} x_4 + a_I \tau_I^2 \varphi(t_{goID}/\tau_I) x_5 - a_D \tau_D^2 \varphi(t_{goID}/\tau_D) x_7 \tag{14}$$

$$\dot{Z}_{IT}(t) = \tau_I \varphi(t_{goIT}/\tau_I) u_I - \tau_T \varphi(t_{goIT}/\tau_T) u_T$$
$$\dot{Z}_{ID}(t) = \tau_I \varphi(t_{goID}/\tau_I) u_I - \tau_D \varphi(t_{goID}/\tau_D) u_D \tag{15}$$

where

$$\varphi(\chi) = e^{-\chi} + \chi - 1 \tag{16}$$

### D. Optimal Pursuit and Evasion Guidance Laws for Interceptor

Let us now present some state-of-the-art linear guidance laws, which are assumed to be used by interceptor

*Lemma 1.* The linear-quadratic optimal guidance law (LQOGL) [3]:

$$u_I^* = \begin{cases} -\dfrac{K(t) Z_{ID}(t)}{\omega_1} u_I^{\max} \tau_I \varphi\left(\dfrac{t_{f2} - t}{\tau_I}\right) & \text{for } \|Z_{ID}(t)\| < \eta \\ -\dfrac{P(t) Z_{IM}(t)}{\xi_1} u_I^{\max} \tau_I \varphi\left(\dfrac{t_{f1} - t}{\tau_I}\right) & \text{else} \end{cases} \tag{17}$$

where $\eta$ is a positive constant representing the limit-collision radius between the interceptor and the defender and $u_I^{\max}$ is the maximum control force provided by the interceptor. Furthermore, variable $K(t)$ and $P(t)$ can be defined as:

$$K(t) = \frac{1}{\int_{t}^{t_{fID}} \left[ \frac{1}{\omega_1} \left( u_I^{max} \tau_I \varphi\left(\frac{t_{fID}-t}{\tau_I}\right) \right)^2 - \frac{1}{\omega_2} \left( u_D^{max} \tau_D \varphi\left(\frac{t_{fID}-t}{\tau_D}\right) \right)^2 \right] dt - 1}$$

(18)

$$P(t) = \frac{1}{\int_{t}^{t_{fIM}} \left[ \frac{1}{\xi_1} \left( u_I^{max} \tau_I \varphi\left(\frac{t_{fIM}-t}{\tau_I}\right) \right)^2 - \frac{1}{\xi_2} \left( u_M^{max} \tau_M \varphi\left(\frac{t_{fIM}-t}{\tau_M}\right) \right)^2 \right] dt - 1}$$

(19)

where $\omega_1$, $\omega_2$, $\xi_1$ and $\xi_2$ are nonnegative constants ensuring the interceptor to converge towards the target, guaranteeing its escape from the defender.

*Proof.* The detailed proof of similar results can be found in [3], see Theorem 1 and the associated proof.

*Lemma 2.* Standard optimal guidance law (SOGL) [20]:

$$u_I^* = u_I^{max} \operatorname{sgn}\left[Z_{ID}(t_{fID})\right] \operatorname{sgn}\left[\varphi\left(\frac{t_{fID}-t}{\tau_I}\right)\right] \quad \text{for } \|Z_{ID}(t)\| < \eta$$

$$u_I^* = -u_I^{max} \operatorname{sgn}\left[Z_{IT}(t_{fIT})\right] \operatorname{sgn}\left[\varphi\left(\frac{t_{fIT}-t}{\tau_I}\right)\right] \quad \text{else}$$

(20)

where $\eta$ is a positive constant representing the switching condition always equal to the defender kill radius.

*Proof.* Consider the following cost function:

$$J_1 = -\frac{1}{2} Z_{ID}^2(t_{fID}) \quad \text{for } \|Z_{ID}(t)\| < \eta$$

$$J_2 = \frac{1}{2} Z_{IT}^2(t_{fIT}) \quad \text{else}$$

(21)

For $J_1$, the Hamiltonian of the problem is defined as:

$$H_1 = \lambda_1 \dot{Z}_{ID}(t)$$

(22)

The costate equation and transversality condition are given by:

$$\dot{\lambda}_1(t) = -\frac{\partial H_1}{\partial Z_{ID}} = 0$$

(23)

$$\lambda_1(t_{fID}) = \frac{\partial J_1}{\partial Z_{ID}(t_{fID})} = -Z_{ID}(t_{fID})$$

(24)

The optimal interceptor controller minimizes the Hamiltonian satisfying the:

$$u_I^* = \arg\min_{u_I}(H_1) \tag{25}$$

The interceptor guidance law can thus be obtained:

$$u_I^* = u_I^{\max} \operatorname{sgn}\left[Z_{ID}(t_{fID})\right] \operatorname{sgn}\left[\varphi\left(\frac{t_{fID} - t}{\tau_I}\right)\right] \tag{26}$$

For $J_2$, similar interceptor guidance law can be found:

$$u_I^* = -u_I^{\max} \operatorname{sgn}\left[Z_{IT}(t_{fIT})\right] \operatorname{sgn}\left[\varphi\left(\frac{t_{fIT} - t}{\tau_I}\right)\right] \tag{27}$$

Finally, the interceptor guidance schemes for evading the defender and pursuing the target are proposed after combining the Eq. (26) and Eq. (27):

$$\begin{aligned} u_I^* &= u_I^{\max} \operatorname{sgn}\left[Z_{ID}(t_{fID})\right] \operatorname{sgn}\left[\varphi\left(\frac{t_{fID} - t}{\tau_I}\right)\right] \quad \text{for } \|Z_{ID}(t)\| < \eta \\ u_I^* &= -u_I^{\max} \operatorname{sgn}\left[Z_{IT}(t_{fIT})\right] \operatorname{sgn}\left[\varphi\left(\frac{t_{fIT} - t}{\tau_I}\right)\right] \quad \text{else} \end{aligned} \tag{28}$$

*Remark 1.* As shown in Eqs. (18) and (19), the LQOGL has to solve the Riccati differential equation. However, its update frequency cannot meet the real-time spacecraft guidance requirements. Compared to the LQOGL, the SOGL in Eq. (28) does not need to solve the Riccati differential equation and has no hyperparameter. This improves both its computation efficiency and robustness at the cost of flexibility and the occurrence of the chattering phenomenon. Because the general family of optimal guidance laws make harsh assumption about perfect information and prior knowledge, they currently do not appear to be implementable in practical use. However, in this work, to demonstrate the performance of the proposed DRL-based guidance law, the target and the defender has imperfect information on the relative states and prior knowledge of the interceptor, while the interceptor is assumed to has perfect information on those of the target and the defender. Considering the above factors, the SOGL was chosen as a benchmark.

## III. Reinforcement Learning-based Guidance Law Development

In this section, the guidance schemes for both the target and the defender partaking in the engagement are developed. We introduce the mathematical formulation of a generic MDP to distinguish perfect and imperfect information situations, which is required to set up the mathematical framework of DRL algorithms. It should be noticed that the game between the interceptor and the alliance of the target and the defender is zero-sum game and herein lies the extremely sparse reward problem [21, 22]. Tricks of reward function design and curriculums induction [23] are presented individually to cope with this issue. According to the existing research, the performance of DRL algorithms is significantly impacted by the implementation. Thus, the choice of algorithms and the implementation details are also demonstrated.

### A. Markov Decision Process

The sequential decision making that an autonomous RL agent interacts with the environment (e.g., the engagement) can be formally described as an MDP, which is required to properly set up the mathematical framework of an DRL problem. A generic time-discrete MDP can be represented as a 7-tuple $\{s, \Omega, o, a, P_{sa}, \gamma, R\}$. $s_t \in S \in \mathbb{R}^n$ is a vector that completely identifies the state of the system (e.g., the EOM) at time $t$. Generally, the complete state is not available to the agent at each time $t$, the decision-making relies on an observation vector $o_t \in O \in \mathbb{R}^m$. In the present paper, the observations are defined as an uncertain (e.g., imperfect and noisy) version of the true state, which can be written as a function $\Omega$ of the current state $s_t$. The action $a \in A \in \mathbb{R}^l$ of the agent is given by a state-feedback policy $\pi : O \to A$, that is, $a_t = \pi(o_t)$. $P_{sa}$ is time-discrete dynamic model describing the transformation led by state-action pair $(s_t, a_t)$. As a result, the evolution rule of the dynamic system can be described as follows:

$$s_{t+1} = P_{sa}(s_t, a_t)$$
$$o_t = \Omega(s_t)$$
$$a_t = \pi(o_t)$$

Since a fixed-time engagement is considered, the interaction between the agent and the environment gives rise to a trajectory $\tau$:

$$s_0, a_0, r_1, s_1, a_1, r_2, \cdots, s_{T-1}, a_{T-1}, r_T, s_T$$

The return, the agent received at time $t$, is defined as a discounted sum of rewards:

$$R_t^\tau = \sum_{i=t}^{T} \gamma^{i-t} r_i$$

where $\gamma \in (0,1]$ is a discount rate determining whether the agent has a long-term vision ($\gamma = 1$) or is short-sighted ($\gamma \ll 1$).

Before deriving the guidance law at hand, key components of the MDP were described: state space, action space, and observations. As the most important part of the setup, the reward design was shown separately.

*1. Perfect Information Model*

In a deterministic model, the baseline assumption is that perfect information of adversary (e.g., relative states, maximum acceleration, and time constant) is available to each of players. Communication of this information between the defender and the protected target is assumed to be ideal and occur with no lag. Thus, the state space can be identified by relative information states and interception time as:

$$s_t = [\boldsymbol{y}_t \quad \boldsymbol{V}_t \quad \boldsymbol{a}_t \quad \boldsymbol{a}_{\max} \quad \boldsymbol{\tau}]^T \tag{29}$$

As for the multi-agent system, interactions cause uncertainty in the environment, significantly impacting the RL algorithm stability. Considering that both the defender and the target are fully cooperative, due to the communication assumption, the model has to learn a guidance law for both the defender and the target. This efficiently reduces the environmental uncertainty and improves the model convergence. In practical application, the same trained agent will be assigned to the target pair. Thus, the action space can be given as follows:

$$\text{action} = [u_T \quad u_D] \tag{30}$$

Since the dynamics of the scenario is formulated in Sec. II A, the state can be propagated implicitly as the linearized equation of motion presented in Eqs. (4) ~ (6).

*2. Imperfect Information Model*

The imperfect information commonly arises due to limitations in radar measurement and deletion of prior knowledge. Nevertheless, in existed study, the perfect information is a strong assumption, which results in implementation difficulties in practice. To address this dilemma, in this work the information degrading is considered. On the one hand, interceptor is assumed to have perfect information (i.e., the relative states and the maneuverability of the target and the defender). On the other hand, the observation of the target and the defender is imperfect and even noise-corrupted. The observation uncertainty is modeled as a measurement noise and a mask on the perfect information

$$o_t = \Omega(s_t) = \boldsymbol{\Gamma} s_t \times (\boldsymbol{I} + \boldsymbol{\omega}_{o,t}) = \begin{bmatrix} \boldsymbol{y}_t \\ \boldsymbol{V}_t \\ \boldsymbol{a}_t \end{bmatrix} \times \begin{bmatrix} \delta \boldsymbol{y}_{o,t} \\ \delta \boldsymbol{V}_{o,t} \\ \delta \boldsymbol{a}_{o,t} \end{bmatrix} \tag{30}$$

being

$$\boldsymbol{\omega}_{o,t} = \begin{bmatrix} \delta \boldsymbol{y}_{o,t} \\ \delta \boldsymbol{V}_{o,t} \\ \delta \boldsymbol{a}_{o,t} \end{bmatrix} \sim U(0_9, \Sigma) \in \mathbb{R}^9 \tag{30}$$

where $\Gamma$ is the mask matrix and $\Sigma = [\sigma_y \boldsymbol{I}_3, \sigma_v \boldsymbol{I}_3, \sigma_a \boldsymbol{I}_3]^T$ represents the noise amplitude, with $\sigma_y$, $\sigma_v$, and $\sigma_a$ the nonnegative parameters.

**B. Reward Function and Curriculum Learning**

Developing a reward function was the most challenging part of solving this multi-agent pursuit-evasion game through RL as the function had to be adaptive to the engagement with a sparse reward setting. It was noticed that, barring the common guidance mission, the pursuit-evasion game can be formulated as a strictly competitive zero-sum game. Moreover, the agent policy network weights were randomly initiated at the beginning of training, while the interceptor was employed with the optimal guidance and is sufficiently aggressive.

In [24], a shaping technique, as a particularly effective approach to solving sparse reward problems, was presented through a set of biological experiments. The researchers broke down a difficult task into several simple units, training the animals following an easy-to-hard schedule. This approach requires adjusting the reward signal to cover the entire training process, followed by gradual modifications to the task dynamics as the training proceeds. In [23], researchers carried forward the idea and proposed curriculum learning, a kind of training strategies. In this work, the shaping technique and the curriculum learning were utilized to speed up the convergence of neural networks and increase algorithm stability and performance.

The goal of the target and the defender is to converge $Z_{ID}$ to zero as $t \to t_{f2}$ while keeping $Z_{IT}$ as large as possible. On the contrary, the interceptor control law is designed to make $Z_{IT}$ converge to zero while maintaining $Z_{ID}$ as large as possible.

For this reason, a non-sparse reward function was defined in Eqs. (31) and (32):

$$r_{medium} = \left|\frac{Z_{IT}}{\alpha_1}\right|^{\beta_1} - \left|\frac{Z_{ID}}{\alpha_2}\right|^{\beta_2} \tag{31}$$

$$r_{terminal} \begin{cases} = \sigma, & \text{if succeed} \\ = -\sigma, & \text{else} \end{cases} \tag{32}$$

where $\alpha_1$, $\alpha_2$, $\beta_1$, $\beta_2$, and $\sigma$ are the positive hyperparameters. It must be stressed that, since both the number and maneuverability of players completely change the environment, the hyperparameter values used in this paper may be not universal. Thus, in the following subsection, the focus will be on the applied design method instead of the specific hyperparameter values. The $r_{terminal}$ is the terminal reward signal given to the terminal behavior of the agent, which is sparse but intuitive. Situations in which the interceptor is destroyed by the defender (when $t = t_{fID}$) or when the interceptor is driven away by the defender and misses the target are judged as a success. Furthermore, the $r_{medium}$ is a non-sparse reward signal achieved using the exponent function, which provides the agent with a smooth reward signal in every state it visits. It carries the physical meaning of the mission – the target must escape from the interceptor, while the defender has to get close to the interceptor. The $r_{medium}$ value increases as $Z_{ID}$ converges to zero, or when $Z_{IT}$ increases. On the other hand, it decreases when $Z_{ID}$ is divergent or when $Z_{IT}$ converges to zero.

Generally, reward normalization is beneficial to neural network convergence. However, determining the bounds of $Z_{IT}$ and $Z_{ID}$ is a complex task. For this reason, hyperparameters $\alpha_1$, $\alpha_2$, $\beta_1$, and $\beta_2$ were tuned, aiming to scale the $r_{medium}$ close to $[-c, c]$, in which $c$ is a positive constant. In the following step, the design of $\rho$ was considered, which introduces the expectation of agent foresight. If the agent is expected to predict the terminal reward $r_{terminal}$ $n$ steps before, the discounted terminal reward must be larger than the $r_{medium}$ bounds. Thus, the hyperparameter $\sigma$ satisfies the following expression:

$$\sigma \geq \frac{c}{\gamma^n} \tag{33}$$

where $\gamma$ is the discount factor in the Markov decision process.

Following with the hyperparameter tuning process, a sequence of increasingly difficult tasks was allocated to the agent, as shown in Table 1. The curriculum was divided into three stages:
- the agent is required to combat the interceptors employing non-maneuvering,

- square wave signal,
- OGL.

Finally, it is possible to complete the reward shaping process.

**Table 1. Curriculum Learning**

| Curriculum | Stage 1 | Stage 2 | Stage 3 |
|---|---|---|---|
| Interceptor guidance command | None | Square wave signal | OGL |
| Maximum interceptor acceleration | 0 | 6g | 2g/4g/6g |
| Episodes | 0~100 | 101~1000 | Remainder |

**C. Reinforcement Learning**

The algorithm adopted in this work is twin delay deep deterministic policy gradient (TD3) [25], which is a model-free, off-policy, actor-critic method widely recognized for the exploration ability and high performance in a series of tough high-dimensional control task. A model-free method is particularly suited for the proposed imperfect information problem in which the transition function $P_{sa}$, observation model $\Omega$, and disturbance distribution cannot be explicitly or analytically expressed, which may be not allowed traditional methods. Also, off-policy approaches gain the advantages of being substantially more sample efficient, because they can reuse history data more efficiently than on-policy method such as PPO [26]. This characterization naturally adapts to the curriculum learning and bring the policy better generalization.

Let us introduce the mathematical framework of the DRL algorithms. Without loss of generality, throughout the entire section the MDP is supposed to be perfectly observable (i.e., with $o_t = s_t$) to conform with the standard notation of RL. However, the perfect information state $s_t$ can be replaced by observation $o_t$ whenever the observations differ from the state.

*1. Actor-Critic Algorithms*

The RL problem goal is to find the optimal policy $\pi_\phi$ with parameters $\phi$ that maximizes the expected return, which can be formulated as follows:

$$J(\phi) = \mathop{\mathbb{E}}_{\tau \sim \pi_\phi}\left[ R_0^\tau \right] = \mathop{\mathbb{E}}_{\tau \sim \pi_\phi}\left[ \sum_{i=0}^{T} \gamma^{i-0} r_i \right]$$

where $\mathbb{E}_{\tau \sim \pi}$ denotes the expectation taken over the trajectory $\tau$. In actor-critic algorithms, the policy, known as the actor, can be updated by using deterministic policy gradient algorithm:

$$\nabla_\phi J(\phi) = \mathbb{E}_{P_{sa}} \left[ \nabla_a Q^\pi(s,a) \big|_{a=\pi(s)} \nabla_\phi \pi_\phi(s) \right]$$

The expected return, when performing action $a$ in state $s$ and following $\pi$ after, is called critic or the value function, which can be formulated as follows:

$$Q^\pi(s,a) = \mathbb{E}_{\tau \sim \pi_\phi} \left[ R_t^\tau | s,a \right]$$

The value function can be learned through off-policy temporal differential learning, an update rule based on Bellman equation which described the relationship between the value of the state-action pair $(s,a)$ and the value of the subsequent state-action pair $(s',a')$:

$$Q^\pi(s,a) = r + \gamma \mathbb{E}_{s',a'} \left[ Q^\pi(s',a') \right]$$

In deep Q-learning[引用], the value function can estimated with a neural network approximator $Q_\theta(s,a)$, with parameters $\theta$ and the network is updated by using temporary differential learning with a secondary frozen target network $Q_{\theta'}(s,a)$ to maintain a fixed objective $U$ over multiple updates:

$$U = r + \gamma Q_{\theta'}(s',a'), \ a' = \pi_{\phi'}(s')$$

where the actions $a'$ are determined by a target actor network $\pi_{\phi'}$. Generally, the loss function and update rule can be formulated as follows:

$$J(\theta) = U - Q_\theta(s,a)$$

$$\nabla_\theta J(\theta) = \left[ U - Q_\theta(s,a) \right] \nabla_\theta Q_\theta(s,a)$$

The parameters of target networks are updated periodically to exactly match the parameters of the corresponding current networks, which is called delayed update. This leads to the original actor-critic method, whose basic structure is shown in Fig. 2.

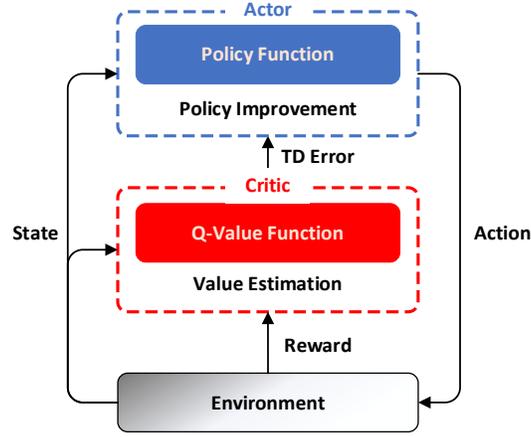

**Fig. 2 Structure of Actor-Critic Method**

*2. Twin Delayed Deep Deterministic Policy Gradient Algorithm*

To address the common RL issues in actor-critic algorithms (i.e., overestimation bias and accumulation of errors), in the TD3 algorithm, the actor-critic framework is modified from three aspects.

- A novel variant of Double Q-learning, called clipped double Q-learning is developed to limit possible overestimation. This gives the update objective of the critic:

$$U = r + \gamma \min_{i=1,2} Q_{\theta'_i}(s', \pi_{\phi'_1}(s'))$$

- Updating the parameters of policy networks periodically to match the value network, which is called delayed policy update, and adopting soft update approach which can be formulated as:

$$\theta' \leftarrow \kappa\theta + (1-\kappa)\theta'$$

where $\kappa$ is a proportion parameter.

- Target policy smoothing regularization is adopted to alleviate the overfitting phenomenon, which can be explicated given as follows:

$$U = r + \gamma Q_{\theta'}(s', \pi_{\phi'}(s') + \varepsilon)$$

where $\varepsilon$ is a clipped Gaussian noise.

An overview of the TD3 algorithm is demonstrated in Fig. 3.

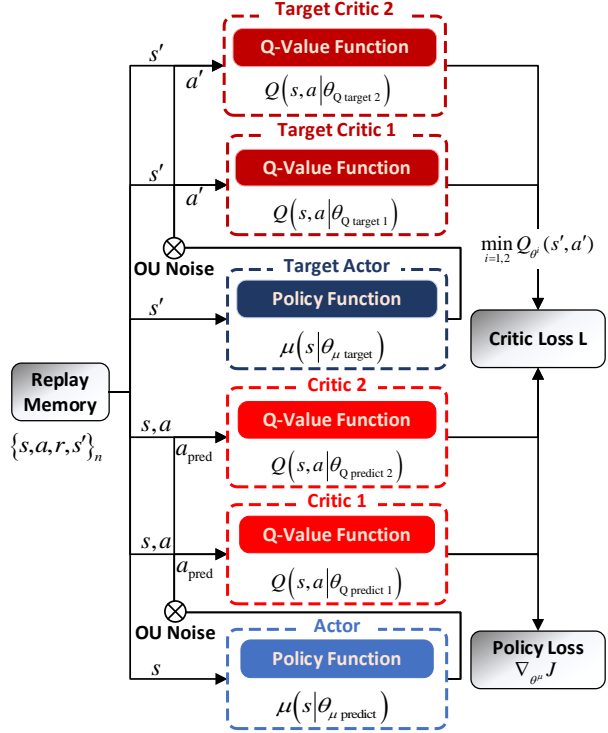

**Fig. 3 Structure of TD3 Algorithm**

*3. Implementation details*

The results presented in this work is obtained by using the TD3 implementation by stable baselines [引用], an open-source library, containing a set of improved implementations of RL algorithm. As for the network architecture setting, the agent observations are vectors with 13 dimensions. Both the guidance policy estimation (actor) and the value function estimation (critic) consist of 3 fully connected layers where the sizes are 64, 256, 512, respectively, along with layer normalization. The output layer has 2 units for the actor, which represent the uniform guidance command of the target and the defender, respectively, and has 1units for the critic. The activation function is Relu for neurons in the hidden layers and is linear for the neuron in the output layer. This structure is devised heuristically and can provably generalize for efficient function approximation. Deeper and wider network is avoided for real-time performance and fast convergence.

The hyperparameters of TD3 have been devised and validated by empirical experiment, which are reported in Table 2.

**Table 2. TD3 hyperparameters**

| Hyperparameter | Symbol | Value |
|---|---|---|
| Discount factor | $\gamma$ | 0.99 |
| Learning rate | $\alpha$ | $3\times 10^{-4}$ |
| Buffer size | $\mathbb{B}$ | 5120 |
| Batch size | $n_{batch}$ | 128 |
| Soft update coefficient | $\tau$ | $5\times 10^{-3}$ |
| Policy delay | $n_{opt}$ | 2 |
| Train frequency | $\omega$ | 6000 |

## IV. Experiments

In this section, the proposed guidance method performance and the shaping technique effectiveness were demonstrated using both the learning process and Monte-Carlo simulation. As an example, consider a scenario consisting of a high agility small spacecraft as the interceptor (I), a defending vehicle as the defender (D), and an evading spacecraft as the target (T). All the players are moving in a circular orbit around the Earth. The gravitational force effect is accounted for in the simulations. It was assumed that the interceptor maneuverability and time constant are superior to those of the target and the defender.

### A. Engagement Setup

In this scenario, the target and the defender altitude are set as $y_T = 500$ km and $y_D = 500.1$ km (corresponding to the reference frame), respectively. To test the intelligent guidance method capacity to adapting environment dynamics, the interceptor altitude was initialized in the range between 499.8 km and 500.2 km. Initial horizontal velocities of the players were set as $V_I = V_T = V_D = 7.6$ km/s. The engagement started when the horizontal distance between the target and the interceptor reached 300 km. The defender was assumed to be launched at the beginning of the engagement, 120 km before the target. Since both the interceptor and the defender aim to intercept their adversaries, they commonly have superior maneuverability compared to the target. This also implies a larger maximum acceleration $u^{max}$ and a smaller time constant $\tau$. In contrast, the target maneuverability is generally limited. Considering all of the above-listed factors, the time constants of each player were set as $\tau_M = \tau_D = 10$ ms, and $\tau_I = 5$ ms. Furthermore, the maximum acceleration of each player was set as $u_I^{max} = 6g$, $u_T^{max} = 1g$, and $u_D^{max} = 2g$. The comprehensive list of engagement parameters is shown in Table 3.

Table 3. Engagement parameters

| Adversary / Parameters | Interceptor | Spacecraft | Defender |
|---|---|---|---|
| | | | |
| Horizontal location | 300 km | 0 km | 120 km |
| Vertical location | 499.8 km ~ 500.2 km | 500 km | 500.1 km |
| Horizontal velocity | - 7.6 km/s | 7.6 km/s | 7.6 km/s |
| Vertical velocity | 0 | 0 | 0 |
| Maximum acceleration | 60 m/s$^2$ | 10 m/s$^2$ | 20 m/s$^2$ |
| Time constant | 0.05 s | 0.1 s | 0.02 s |
| Kill radius | 0.25 m | 0.5 m | 0.25 m |

**B. Experiment 1: Convergence and Real-Time Performance of the Guidance Policy**

To verify that the proposed scheme (i.e., non-sparse reward with CL) can accelerate convergence and acquire higher accumulated reward, the learning processes, using sparse/non-sparse reward signal and non-sparse reward with CL, respectively, with same hyperparameters, were demonstrated. During the learning process, weights of the neural network model were saved every 100 episodes for subsequent analysis. Furthermore, to remove stochasticity as a confounder, six random seeds were set for each case. Meanwhile, the real-time performance of the optimized agent is evaluated by comparing with the traditional OGLs.

The agents were obtained after a training consisting 20000 episodes, which took 12 hours with 8 parallel workers on a computer equipped with 104-core Intel Core Xeon Platinum 8270 CPU @2.70 GHz. Similarly, both the traditional methods and the proposed method are given a current state or observation, and return the required action. Table 4 exhibits the comparison of the calculation cost and update frequency, obtained by using SOGL, LQOGL and the proposed method. It can be seen from the table that the LQOGL is time consuming due to the calculation of Riccati function, which is the reason that it has not been practically applied. As a proven approach, the SOGL has an outstanding real-time performance. The proposed method can achieve an update frequency of $1.1 \times 10^3$ HZ Hz and shows great potential of on-board applications. Whereas, a variety of approaches (e.g., pruning and distilling) are effective to compress the policy network and further improve its real-time performance, it's not the main work of this research.

**Table 4. Statistic of time consumption with different guidance method**

| Metrics | LQOGL | SOGL | RL-base guidance |
|---|---|---|---|
| Duration (1e3 Step) | 2.773 s | 0.0145 s | 0.910 s |
| Update Frequency | $\approx 360$ HZ | $\approx 6.9 \times 10^4$ HZ | $\approx 1.1 \times 10^3$ HZ |

Additionally, the learning curves, presenting the mean accumulated rewards vs. learning episodes, of various cases are shown in Fig. 3. As illustrated in Fig. 3, under the sparse reward setting, the accumulated reward stayed $-\rho$ due to the rewards that mark progress toward the goal is infrequent. A relatively steady improvement was noticed in accumulated reward when the agent employed either a non-sparse reward signal or a non-sparse reward signal with the addition of CL. The agent learning with non-sparse reward signal uses around 6000 episodes to achieve the convergent situation. More notably, the agent employing non-sparse reward signal with CL takes only 1800 episodes after the curriculum (i.e., 4300 episodes at all) to exceed the agent without CL and finally achieves higher convergent accumulated reward. It can be conclude that the proposed reward function and learning schemes organize exploration properly and solve the sparse reward problem and hence exhibit superior learning efficiency and asymptotic performance.

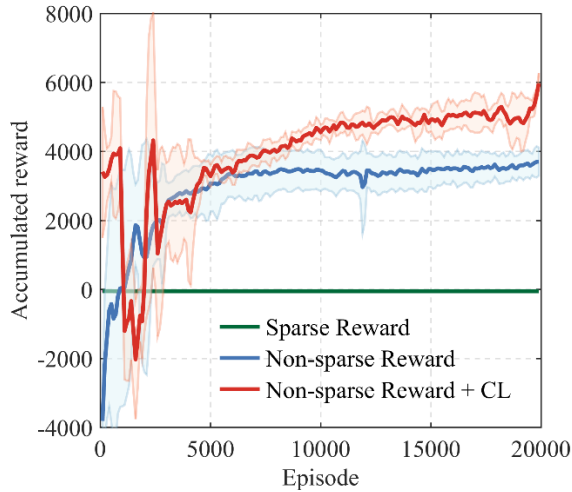

Fig. 3 Learning curves of the TD3

## C. Experiment 2: Validation in Fully Observable Multi-Agent Game

The performance of the trained agent in the fully observable game is investigated, by comparing the win rate corresponding to an optimized policy $\pi_\phi(s)$, obtained by performing Monte-Carlo simulation in the fully observable (deterministic and with default engagement parameters) environment, with the solution of the SOGL.

*1. Baselines*

The SOGL for the target and the defender were considered as a benchmark. Through brief derivation similar to that in Section III, it can be proven that the SOGLs for the target and the defender are:

$$u_T = -u_T^{max} \text{sgn}\left[Z_{IT}\left(t_{fIT}\right)\right]\text{sgn}\left[\varphi\left(\frac{t_{fIT}-t}{\tau_T}\right)\right]$$

$$u_D = u_D^{max} \text{sgn}\left[Z_{ID}\left(t_{fID}\right)\right]\text{sgn}\left[\varphi\left(\frac{t_{fID}-t}{\tau_D}\right)\right] \tag{34}$$

*2. Win Rate*

Fig. 4 shows the curves that represent win rates vs. learning episodes, corresponding to the agents learning with and without CL. The bright-yellow line indicates that the SOGL has a win rate of 87.0% in a deterministic environment. The agent learning without curriculum uses 3200 episodes exceeding the baseline and uses around 9500 episodes to reach the best performance of 99% win rate. However, the performance gradually drops after reaching the top and finally converges at around 70%, which indicates that overfitting occurs when the policy network continues the training. When employing the curriculums defined in Sec. II, it takes around 6400 episodes for the agent to achieve the convergent situation of 99% win rate. Faster training process convergence speed and better local optimum values were detected in the agent learning with curriculum. Moreover, the proposed methodology has shown capability in mitigating the overfitting phenomenon.

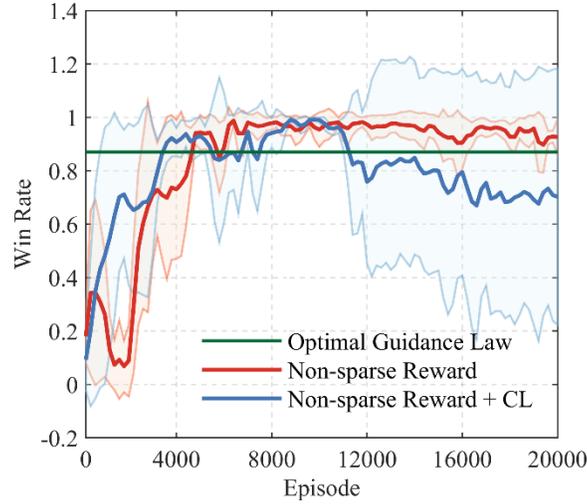

Fig. 4 Win Rate

*3. Performance Test*

Fig. 5 presents a sample run of the trajectories of the three players, while Fig. 6 and Fig. 7 present their guidance command and the ZEMs between the target and the interceptor and between the defender and the interceptor, respectively. In this scenario the interceptor is initialized with three typical positions.

The cooperation between the target and the defender, established by capitalizing the relative states information, is evident. As Fig. 5 depicted, the defender is able to expel the interceptor that incoming with different impact angle and guarantee the successfully evading of the target, which illustrated that the proposed guidance strategy is shown to be adaptive over a wide range of incoming direction. As shown in Figs. 6 and 7, the target constantly changes maneuver, acting like a bait to lure the interceptor into the kill radius of the defender and maximizing the target-interceptor ZEM in the meantime.

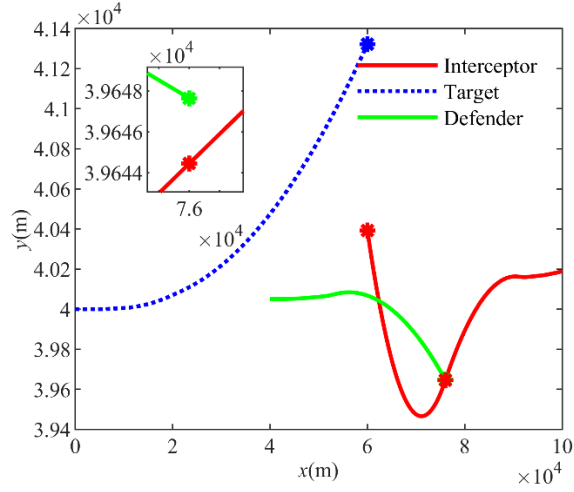

Fig. 5 Trajectory of case 1, 2, and 3

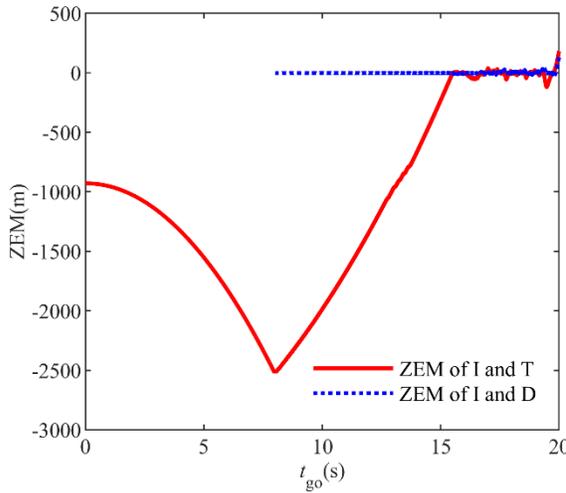

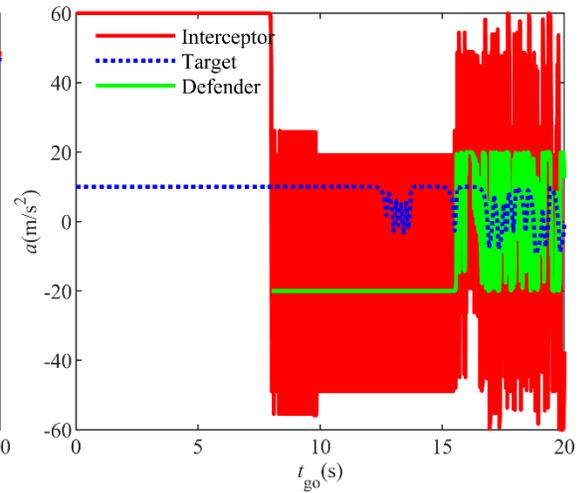

Fig. 6 ZEM of case 1　　　　　　　　　　Fig. 7 Guidance command of case 1

**D. Experiment 3: Adaptiveness of the RL-based Guidance Method**

To validate that the optimized policy can adapt to different engagement parameters, we design a comparative experience where the agent, trained in the fixed environment formulated in this section, is deployed to combat with interceptors with different maneuverability. The relative performance of the proposed method, along with the SOGL, were investigated.

Fig. 8. shows the win rates, correspond to different interceptor maximum accelerations and time constants. It is evident from Fig. 8 (a) that for both guidance strategies, the performance declines as the maximum interceptor acceleration increases. When combatting the interceptor with a maximum acceleration not more than that in training environment, the agent performance is better. However, when combatting the interceptor whose maximum acceleration above 6 g, the agent performance is not good enough. Thus, it can be concluded that the adaptiveness of the trained agent is limited to the maximum interceptor acceleration.

On the other hand, the variation on the interceptor time constant has little effect on the proposed guidance method, while the SOGL are sensitive to it. As shown in Fig. 8 (b), the win rate of the SOGL drops sharply as the interceptor time constant decreases i.e., as its agility improves.

**Remark 3.** As an analytical method, the SOGL is stable but inflexible, which is incurred by its theoretical framework [27] along with harsh assumptions [28]. Correspondingly, the RL-based guidance strategies are flexible and can be optimized continuously. The proposed method is independent of the time constant, meaning that it perform better with less prior knowledge than the SOGL. Moreover, the adaptiveness of the proposed method can be improved by considering the tolerance of maximum interceptor acceleration.

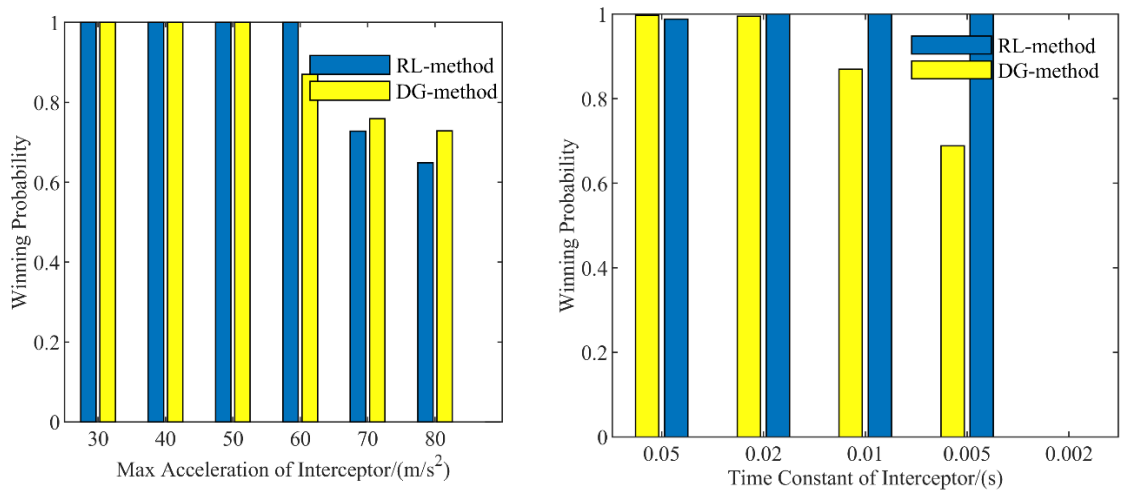

Fig.8 (a) (b) Win rate in situations without priori knowledge

**E. Experiment 4: Robustness of the RL-based Guidance Method**

Besides the unperturbed, fully observable game, the following noise-corrupted, partially observable game studies have been analyzed separately in this manuscript. The parameters used for describing the imperfect information model defined in Sec. II is presented in Table 5. The win rates of the proposed method and the SOGL that results from the Monte-Carlo simulation are shown in Fig. 9.

**Table 5. Parameters of the different imperfect information model**

| Measurement Noise | Parameter | Case 1 | Case 2 | Case 3 | Case 4 | Case 5 | Case 6 |
|---|---|---|---|---|---|---|---|
| Bearing | $\sigma_y$ | 0 | 0 | 0 | 0.01 | 0.02 | 0.05 |
| Velocity | $\sigma_v$ | 0.05 | 0.1 | 0.2 | 0.05~0.45 | | |
| Acceleration | $\sigma_a$ | | 1~3 | | 1 | 1 | 1 |

First, it's worth noticing in Fig. 9 (a) ~ (c) that when the bearing can be estimated accurately (i.e., $\sigma_y = 0$), the performance of the policy is insensitive to noise (declined less than 25%) in the velocity and acceleration measurement; hence, the robustness of the proposed method is superior to that of the analytical method over all the scenario. An interesting situation is noticed in Fig. 9 (d) ~ (f) that even though the policy, affected by noise in bearing measurement, becomes sensitive to noise in velocity measurement, the performance is not declined as the noise in bearing measurement enhanced. The reason is probably that the policy network propagates observation information with different weights, which is benefit from the exploration mechanism of RL. Thus, training the agent in the deterministic environment leads to a robust guidance policy with strong noise-resistant capability.

## V. Conclusion

A cooperative guidance strategy was given for active-defense spacecraft, developed using the deep reinforcement learning algorithm. A neural network-based agent was trained for a multi-agent pursuit-evasion engagement. The benefits of the proposed guidance strategy, along with its effectiveness, were verified through the Monte-Carlo simulation.

The second contribution of this work is the development of a universal reward function design approach and the increasingly difficult learning approach. Both were found to significantly accelerate the training process convergence, alleviating the overfitting problem. The developed approach is flexible and can be easily extended to similar multi-agent games. The example includes two-on-two engagements.

The simulation results have shown that the proposed guidance strategy can guarantee the convergence of relative distance between the interceptor and the defender, as well as the target evasion. The guidance command generation is independent of the player maximum acceleration and the time constant. Compared to the SOGL, the deep reinforcement learning-based strategy has improved performance while requiring less prior knowledge.

Furthermore, the proposed guidance strategy was shown to be adaptive to interceptor maneuverability uncertainty, including the maximum acceleration, time constant, and initial position.

As a neural network-based strategy, the computational requirement to generate the guidance command was modest. Thus, real-time applicability could be achieved.

Some future research directions deserve further attentions. For example, it would be worthwhile to explore the adaptiveness of the proposed guidance strategy under spacecraft sensor noise and variable environment dynamics. Efforts can also be made to reduce fuel cost and alleviate saturation and chattering phenomena.

## Acknowledgments

The work described in this paper was supported by the National Natural Science Foundation of China (Grant No. 62003375). The authors fully appreciate their financial supports.